\documentclass[a4paper,12pt]{article}

\usepackage[utf8]{inputenc}
\usepackage[OT1]{fontenc}

\usepackage[square,numbers]{natbib}
\usepackage{graphicx}
\usepackage{epsfig}

\usepackage{latexsym}
\usepackage{array}
\usepackage{amsmath}
\usepackage{amsfonts}
\usepackage{amssymb}
\usepackage{dsfont}

\usepackage{setspace}
\usepackage{float}
\usepackage{textcomp}
\usepackage{color}
\usepackage{url}
\usepackage{times}

\newcommand{\mb}{\mathbf}

\newcommand{\beq}{\begin{equation}}
\newcommand{\eeq}{\end{equation}}
\newcommand{\beqna}{\begin{eqnarray}}
\newcommand{\eeqna}{\end{eqnarray}}
\newcommand{\bit}{\begin{itemize}}
\newcommand{\eit}{\end{itemize}}

\begin{document}
\title{On the use of Singular Spectrum Analysis}
\author{A.M. Tom\'{e}, D. Malafaia, A.R. Teixeira, E.W. Lang}
\maketitle


\begin{abstract}
Singular Spectrum Analysis (SSA) or Singular Value Decomposition (SVD) are often used to de-noise univariate time series or to study their spectral profile. Both techniques rely on the eigendecomposition of the correlation matrix estimated after embedding the signal into its delayed coordinates.
In this work we show that the eigenvectors can be used to calculate the coefficients of a set of filters which form a filter bank. The properties of these filters are derived. In particular we show that their outputs can be grouped according to their frequency response. Furthermore,  the frequency at the maximum of each frequency response and the corresponding eigenvalue can provide a power spectrum estimation of the time series. Two different applications illustrate how both characteristics can be applied to analyze wideband signals in  order to achieve narrow-band signals or to infer their frequency occupation.
\end{abstract}
\section{Introduction}
Projective subspace models, applied to time series data sets, can be found in literature under various names depending on the domain of application: Singular Spectrum Analysis (SSA) (for instance in climate time series analysis) \citep{Golyandina01}, \citep{Ghil02}, \citep{Teixeira06b} or Singular Value Decomposition (SVD) (for instance in speech enhancement \citep{Ephraim95,Hansen07,Hermus07,Hassanpour2012} and more recently in spectrum sensing \citep{Axell2012,Huang2014}). The aim of SSA is to decompose a time series into a sum of a small number of interpretable components such as a slowly varying trend, oscillatory components and superimposed noise interferences. Applied to speech enhancement, SSA simply serves to eliminate noise which is usually considered additive, independent and normally distributed. In spectrum sensing applications, the eigenvalue profile is used to decide wether a communication channel is occupied or not, or simply to estimate the noise floor of the channel. Thus spectrum sensing applications
are concerned mainly with the study of the spectral profile of the signals. Recently, cognitive radio applications give another perspective to the application of these singular-value-based decomposition methods. In general, the issue is to detect the presence of a signal (user) in a given narrowband channel \citep{Zeng2009}.

SSA also has been applied to process biomedical signals with different goals: eliminating high-amplitude artifacts \citep{Sanctis11}, suppressing noise contributions or extracting informative components \citep{Sanei2011, Ghaderi2011}. The elimination of high-amplitude artifacts is naturally related to the components associated with the largest eigenvalues. Therefore, the number of eigenvalues is always discussed but in most of the applications eliminating the component related with the largest eigenvalue reduces significantly the artifact-related interference without distorting the underlying signal \citep{Sanctis11, Santos2016}. On the contrary, reducing white noise interferences usually is addressed by eliminating components related with small eigenvalues \citep{Ghaderi2011}.  The extraction of components and their grouping according to some criterion has been addressed also by recognizing different trends on the eigenvalue spectra \citep{Ghaderi2011}, or based on certain statistics of the extracted components \citep{
Sanei2011}. Note that in most biomedical applications, as in many other fields in science as well, the power spectra show a $1/f$ dependence in the low frequency range which is matched by an eigenvalue distribution arranged in decreasing order of magnitude \citep{Li2012}.

But in a more general perspective, such subspace models allow to decompose a time series into several components which reflect distinct frequency bands of the original signal.  In general, SSA \citep{Golyandina01} - or SVD \citep{Ephraim95} - based  subspace approaches can be summarized with the following steps :

\begin{enumerate}
\item
Embedding the time series into its time-delayed coordinates resulting in a sequence of lagged multidimensional data vectors. The latter are arranged as columns of a {\em trajectory matrix} with either {\em Toeplitz} or {\em Hankel} structure.
\item
Estimating an orthogonal basis vector matrix, using singular value decomposition or principal component analysis (PCA).
\item
Projecting the multidimensional data vectors onto the new basis vectors.
\item
Selecting {\em relevant} components.
\item
Reconstructing the multidimensional embedded data using a generally lower-dimensional subspace representation.
\item
Employing diagonal (or anti-diagonal) averaging to reconstitute the {\em Toeplitz} or {\em Hankel} structure of the reconstructed trajectory matrix.
\item
Reverting the embedding to yield a univariate time series.
\end{enumerate}

Thus, the SSA - or SVD - based subspace model corresponds to an orthogonal matrix whose column vectors form an eigenbasis of the multidimensional space created by the embedding. {\em Relevant} components of the signal can be obtained by projecting the data onto that eigenbasis and omitting {\em irrelevant}, for example noise -related, components. Finally an improved univariate time series can be reconstructed.

The selection of {\em relevant} components is a critical issue that needs to be addressed each time a problem-specific subspace model is generated. For instance, in \citep{Alexandrov08, Alexandrov09}, the frequency profile of the basis vectors was studied to select those vectors (and related components) which correspond to the trends in the data. In \citep{Tome10b} it is shown that SSA can equivalently be formulated as a parallel filter bank structure where the finite impulse response (FIR) filters represent the eigenvectors of the correlation matrix. During the analysis step, the filter bank decomposes the input signal into $M$ components using the eigenvectors as causal (or anti-causal) FIR filters. Afterwards, the multidimensional reconstruction, including the subsequent diagonal averaging, is correspondingly explained as an anti-causal (causal) FIR filter. The latter filter coefficients simply correspond to the eigenvector entries written in reversed order. The interpretation of a signal enhancement as a linear filtering operation  was discussed in \citep{Hansen98, Hansen07}, and in \citep{Tome10b}, the frequency response of the filters is deduced in closed form. Using techniques from linear time-invariant systems theory to compute input - output relationships \citep{Jackson91, Jackson96}, a closed-form expression for the frequency response of the filters is obtained such that the frequency profile of the output can be extracted easily.  Two important issues are relevant in SSA: the value of the embedding dimension, which is an user predefined parameter,  and the choice of relevant components.

In this work we show that a study of the frequency response of the equivalent FIR filters favorably complements a study of the eigenvalue spectrum and a related choice of the number of {\em relevant} eigenvalues \citep{Alharbi2016}. It is also shown that the eigenvalue profile of the input signal is related to its power spectrum if the eigenvalues are ordered according to the maximal value in the absolute value of the frequency response of the corresponding filter. We illustrate the application of these ideas in two typical applications where such methods may be employed advantageously.

The first example concerns a biomedical application, more specifically an {\em Electroencephalogram} (EEG). The latter represents recordings of electrical brain activities via several sensors thus creating a multichannel biomedical signal that is often used for medical diagnosis or in brain studies. In both cases, artifacts frequently turn the visual analysis difficult and often even impossible. Therefore, pre-processing techniques like digital filtering are often applied to extract the characteristic bands of the EEG signals: delta ($\delta$), theta ($\theta$), alpha ($\alpha$), beta ($\beta$) and  gamma ($\gamma$). Moreover, as nowadays the acquisition boards work with high sampling rates, the recorded input signals have a bandwidth which is larger than what is really useful in most of these studies. Hence, as is shown in this study, extracting only the problem-related informative components of the signals represents a pre-processing technique that can be used profitably in EEG analysis.

The second application concerns a telecommunication application. {\em Cognitive Radio} (CR) is a recent technique the goal of which is to make use of at times unoccupied Radio Frequency (RF) bands through intelligent spectrum sensing \citep{Idrees2015}. Therefore, the study of spectrum analysis methods regained interest, especially studying methods which are able to cope with very low signal-to-noise ratios. These algorithms need to be able to accurately detect the presence or absence of an active user. To detect the occupation of a channel, usually a decision threshold is defined based on the estimation of the noise-floor. One of the available methods to detect the presence of an active user is based on the dynamic range of the eigenvalue spectrum, i.~e. the ratio between the maximal and minimal eigenvalues of the correlation matrix of the sensed signal \citep{Zeng2009}.

The paper is organized as follows: in section 2 SSA and its representation as a filter bank is presented before in section 3 some illustrative applications are discussed and finally, in section 4, some conclusions are drawn.

\section{ Singular Spectrum Analysis}
\label{sec2}

Time series analysis techniques often rely on embedding one-dimensional sensor signals in the space of their time-delayed coordinates. Embedding can be regarded as a mapping that transforms a one-dimensional time series into a multidimensional sequence of lagged vectors. Considering an univariate signal
\[
(x[0], x[1], \ldots ,x[N-1])
 \]
 with $N$ samples, its  multidimensional variant is obtained by $\mathbf{x}_k = (x[k-1+M-1], \ldots , x[k-1])^\mathrm{T} ,  k=1, \ldots , K $, where $K=N-M+1$. These lagged vectors form the columns of the related data matrix $\mathbf{X}$ which is called a {\em trajectory matrix} \citep{Golyandina01}. The column vectors $\mathbf{x}_k$ of $\mathbf{X}$ span an embedding space of dimension $M$.

\begin{equation}
\small{
\mathbf{X} = \left[
\begin{array}{cccc}
x[M-1] & x[M]   & \ldots & x[N-1] \\
x[M-2] & x[M-1] & \ldots & x[N-2] \\
x[M-3] & x[M-2] & \ldots & x[N-3] \\
\vdots   & \vdots   & \ddots & \vdots  \\
x[0]   & x[1]   & \ldots & x[N-M]
\end{array}
\right].
\label{eqtraj}
}
\end{equation}

\noindent Note that the trajectory matrix has identical entries along its diagonals. Such a matrix is called a {\em Toeplitz matrix}. There are other alternatives to organize a data matrix via embedding the univariate time series signal in an $M$- dimensional space of its time-delayed coordinates. For example, an {\em Hankel matrix} is obtained if the embedding is arranged such that identical elements occur along the anti-diagonals \citep{Golyandina01, Hansen07}. Anyway, by computing two-point time correlations between the entries of the multi-dimensional representation of the signal, it is possible to find an orthogonal matrix $\mathbf{U}$ whose columns form an orthogonal basis of the $M$-dimensional space. The non-normalized $M \times M$ - dimensional correlation matrix is obtained as

\[
\mathbf{R} = \mathbf{X}\mathbf{X}^\mathrm{T}
\label{eq-cor}
\]

\noindent and its eigenvalue decomposition

\[
\mathbf{R} = \mathbf{U}\mathbf{\Lambda}\mathbf{U}^\mathrm{T}
\]
\noindent provides the related subspace model via the matrix $\mathbf{U}$ of basis vectors and corresponding eigenvalues $\mathbf{\Lambda} = \mathrm{diag}(\lambda_1, \cdots ,\lambda_M)$.

Furthermore, the subspace model allows to re-write the original data matrix into a sum of rank-one matrices according to

\begin{eqnarray}
\hat{\mathbf{X}} &=& \mathbf{u}_1 p_{11} \mathbf{u}_1^\mathrm{T} \mathbf{X} + \mathbf{u}_2 p_{22} \mathbf{u}_2^\mathrm{T} \mathbf{X} + \ldots + \mathbf{u}_M p_{MM} \mathbf{u}_M^\mathrm{T} \mathbf{X} \nonumber \\
&=& \mathbf{u}_1  p_{11} \mathbf{Y}_1 + \mathbf{u}_2  p_{22} \mathbf{Y}_2 + \ldots + \mathbf{u}_M   p_{MM}\mathbf{Y}_M \nonumber \\
&=& \sum_{m=1}^M \mathbf{\hat{X}}_m,
\label{eq-rec2}
\end{eqnarray}

\noindent where the $p_{mm}$ denote proper weights, and each element of the $m$-th $1 \times K$ - dimensional matrix $\mathbf{Y}_m = \mathbf{u}_m^T \mathbf{X}$ contains the dot product between the $m$-th eigenvector and the $K$ columns of the data matrix. These projections have the following properties:

\begin{itemize}
\item They are uncorrelated, i.~e. $\mathbf{Y}_m\mathbf{Y}_k^\mathrm{T}=0, k\neq m$
\item Their energy is equal to $\lambda_m$ :$\mathbf{Y}_m\mathbf{Y}_m^\mathrm{T}=\lambda_m$
\end{itemize}

These properties can be easily verified by considering an SVD of the original trajectory matrix according to $\mathbf{X}=\mathbf{U}\mathbf{\Lambda}^{1/2}\mathbf{V}^\mathrm{T}$. Here, the $K \times M$ - dimensional matrix $\mathbf{V}$ has also orthogonal columns as it corresponds to the right eigenvector matrix of $\mathbf{X}$. Therefore, the above discussed projections $\mathbf{Y}_m$ can be expressed as $\mathbf{Y}_m=\lambda_m^{1/2}\mathbf{v}_m^\mathrm{T}$, i.~e. with the right eigenvector scaled by the related singular value.

The most straightforward and indeed widely used technique to select the projections is to consider binary weights, i.~e. $p_{mm} = 1$, if the $m$-th projection is selected for the reconstruction, otherwise $p_{mm}=0$. But in noise reduction applications, for example, the selected {\em relevant}  projections, corresponding to the $L \le M$ largest eigenvalues, are scaled by appropriate normalized weights to reduce the noise contribution. This way, the eigenspectrum of $\hat{\mathbf{X}}$ is a truncated and re-scaled version of the eigenspectrum of the original data. Therefore, assigning a real value $p_{mm} \in \mathds{R}$ like, for instance, $0 \le p_{mm} = \sqrt{1 - \eta/\lambda_m} \le 1$ where $\lambda_m$ denotes the variance of the $m$-th eigenvector and $\eta$ represents the noise variance (or energy), yields for the weighted or re-scaled projections:

\begin{equation}
p_{mm} \mathbf{Y}_m = p_{mm}\mathbf{u}_m^\mathrm{T}\mathbf{X} = (\lambda_m-\eta)^{1/2}\mathbf{v}_m^\mathrm{T}
\end{equation}

Note that only $L \le M$ eigenvectors and related eigenvalues are retained onto which the data is projected. Rescaling thus effectively subtracts out the effect of noise contributions. Thereby, the  noise variance of the data can be estimated as the mean of the $M - L$ discarded eigenvalues  \citep{Liavas01}, but more sophisticated schemes have been considered also. Though other possibilities to compute the weights $p_{mm}$ can be found, all assume that the noise variance should be smaller than the selected {\em relevant} eigenvalues \citep{Hansen07, Hermus07}. In the case discussed here this is assured as $\lambda_m$ only refers to the eigenvalues which are kept, and eigenvalues are ordered with decreasing value.

Concerning the reconstructed data matrix $\hat{\mathbf{X}}$ as well as its rank-one components $\hat{\mathbf{X}}_m$, they generally do not exhibit identical elements along each descending diagonal like the original trajectory matrix $\mathbf{X}$. In SSA or SVD applications, these distinct entries along each diagonal (or anti-diagonal) are replaced by their mean to reconstitute a Toeplitz or a Hankel matrix. Finally, a reconstructed univariate time series $\hat{x}_m[n]$ is obtained by reverting the embedding, i.e. by building the time series, i.~e. the signal, from the mean values of each diagonal (or anti-diagonal) of $\hat{\mathbf{X}}_m$.

After having exposed the main properties of an SSA analysis, in the following we provide a comparative exposition of a classical time domain SSA analysis and an equivalent filter bank description to highlight the complementary information one can gain from both analysis techniques.

\begin{figure}[!htb]
\begin{center}
  \includegraphics[width=0.9 \textwidth]{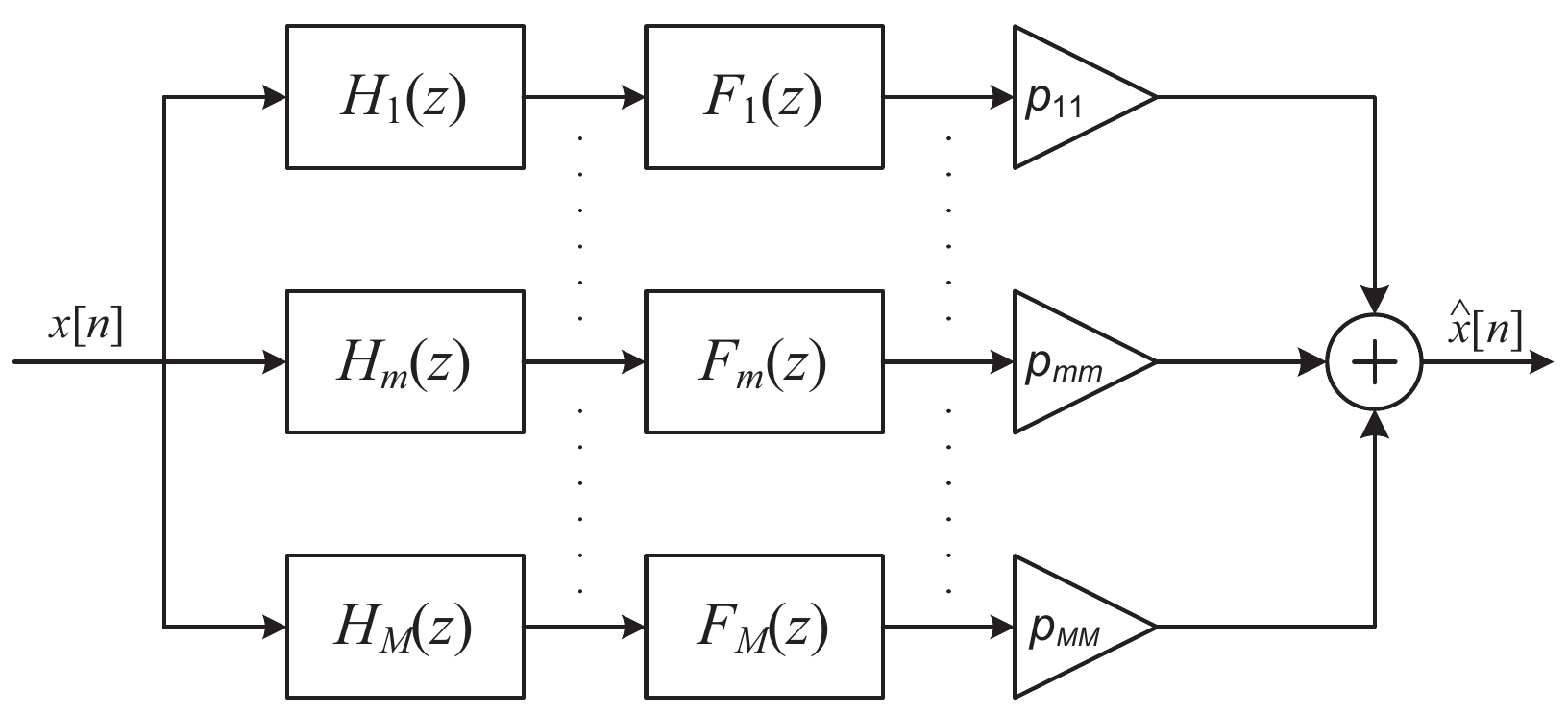}
\end{center}
   \caption{Filter bank description of the SSA processing chain: Analysis ($H_m(z)$) and synthesis ($F_m(z)$) transfer functions and the scaling factor.}
   \label{figblock}
\end{figure}
\subsection{SSA components: time domain}
\label{sec4}

Several works have addressed SSA - or SVD - based subspace projection techniques using linear filtering theory \citep{Hansen98, Dologlou91,Alexandrov09, Alexandrov08, Tome10b}. Concerning signal enhancement applications, in \citep{Hansen98} the manipulations of the input signal were described as a filter bank approach, but all operations are justified using matrix manipulations as suggested by \citep{Dologlou91}. Examples of frequency responses of the eigenfilters, i.~e. the eigenvectors, are shown graphically in \citep{Hansen98}, but no analytic expressions of the filter responses were given. In \citep{Alexandrov08}, the frequency profile of the eigenvectors was studied by way of the corresponding periodogram. In \citep{Tome10b}, an approach based on linear invariant system theory was presented to compute transfer functions of the discrete systems related with the projection step as well as the transfer functions related with the reconstruction and diagonal averaging steps.  The coefficients of both filters
correspond to the components of the related eigenvectors and define a causal and anti-causal filter, respectively. The transfer function of the cascaded causal and anti-causal filters, the $m-th$ branch of the filter bank, is given by

 \begin{eqnarray}
 T_m(z) &=& \frac{\hat{X}_m (z)}{X(z)} = H_m(z)F_m(z)  \nonumber \\
 &=& \left( \sum_{k=0}^{M-1} u_{km}z^{-k} \right)\left( \frac{1}{M} \sum_{k=0}^{M-1} u_{km}z^{k} \right)
 \end{eqnarray}

 \noindent where the filter coefficients $u_{km}$ are the entries of the corresponding eigenvector
 \[
  \mathbf{u}_m = \left( \begin{array}{ccccc}
                   u_{0m} & u_{1m} & u_{2m} &  \ldots & u_{(M-1)m}
                   \end{array}\right)^T.
 \]
Fig. \ref{figblock} illustrates the block diagram of the system, where the scaling coefficient $p_{mm}$ modifies the contribution of each signal component $\hat{x}_m[n]$ to the output $\hat{x}[n]$.

The analysis filter $H_m(z)$ is described by a polynomial with negative exponents (causal), while for the synthesis filter $F_m(z)$, the polynomial has positive exponents (anti-causal) and the coefficients are divided by $M$ due to the diagonal averaging. The product of the two polynomials leads to the transfer function of an equivalent system which can be written as

\begin{equation}
T_m(z) = \frac{\hat{X}_m (z)}{X(z)} = F_m(z)H_m(z) = \sum_{k=-(M-1)}^{M-1} t_{km} z^{k}.
\label{eq_fil}
\end{equation}

The coefficients $t_{km}$ of this system of cascaded analysis-synthesis filters have special properties which can be derived easily from the SSA multidimensional approach. These coefficients are the result of the sum of the diagonals of $\mathbf{T}_m = \mathbf{u}_m \mathbf{u}_m^\mathrm{T}$, e.g, the $k$ - th  element of $t_{km}$ is the sum of the entries along the $k$ - th diagonal of $\mathbf{T}_m$ divided by $M$. Here, the main diagonal corresponds to $k=0$ and positive (negative) values correspond to diagonals above (below) the main diagonal. Therefore, the properties result from the following considerations:

\begin{itemize}
\item

The matrix $\mathbf{T}_m$ is symmetric, i.~e. after adding and dividing  each diagonal by $M$,

\[
t_{|k|m}= \frac{1}{M} \sum \limits_{i=|k|}^{M-1}u_{im}u_{(|k|-i)m}
\]

the coefficients of the filter obey to the relation

\begin{equation}
t_{km} = t_{-km}, \; k=1, \ldots ,(M-1)
\label{eq_sym}
\end{equation}

\item
The component matrices sum up as follows: $\sum \limits_{m=1}^M \mathbf{T}_m = \mathbf{u}_m \mathbf{u}_m^\mathrm{T} = \mathbf{I}$, where $\mathbf{I}$ is the identity matrix. Therefore, the sum of the impulse responses of the $M$ filters yields the unit impulse signal:

\begin{equation}
\sum_{m=1}^M t_{km} = \delta[k]
\label{eq_soma}
\end{equation}

\end{itemize}

This last property of the coefficients of the combined analysis - synthesis filter cascade implies that by applying the filters $t_{km}, m=1 \ldots M$ with maximal weight $p_{mm}=1$ when decomposing any input signal into $M$ in-phase components, at the output the input signal $x[n]$ is recovered. Thus the output signal $\hat{x}[n]$, obtained by adding up filter outputs, is equal to the input signal $x[n]$. The symmetry of the coefficients leads to real-valued frequency responses which can be obtained by substituting $z = e^{j \omega},\ j = \sqrt{-1}$ in \ref{eq_fil}, which then reads

\begin{equation}
T_m(e^{j\omega}) = t_{0m} + \sum_{k=1}^{M-1} 2 t_{km}\cos (k \omega ).
\label{eq_filresp}
\end{equation}

Note that the real-valued frequency response with normalized sampling rate, i.~e. period $\omega = 2\pi$, corresponds to a zero-phase filter. This means that each extracted component $\hat{x}_m[n]$ is always in-phase with its related input signal $x[n]$.  Note that frequency responses of the component filters $H_m(z)$ and $F_m(z)$ cannot be given in closed-form similar to $T_m(z)$ in eqn. \ref{eq_filresp} due to lacking symmetry properties of their coefficients \citep{Jackson96}.

In \citep{Harris2010}  the correlation matrix  $\mathbf{R}$ was forced to form a symmetric Toeplitz matrix which then has symmetric or anti-symmetric eigenvectors. In that case, the frequency responses of the analysis and synthesis filters  can be expressed as $A(\omega)\Phi (\omega)$ where the $A(\omega)$ represents a real-valued function related with the magnitude of the frequency responses,  and $\Phi(\omega)$ represents the phase \citep{Jackson96}.  However, the resulting $T_m(z)$ also follows the same characteristics as described in eqn. \ref{eq_sym}. This is because the same procedure applies to its calculation, and its final expression corresponds to eqn. \ref{eq_filresp}

Naturally, eqn. \ref{eq_fil} can be similarly expressed in the time domain, leading to the implementation equation

\begin{equation}
\hat{x}_m[n]= \sum_{k=0}^{M-1} t_{km} x[n-k] + \sum_{k=1}^{M-1} t_{km} x[n+k]
\label{eq_imp}
\end{equation}

which corresponds to the time domain convolution of the non-causal filtering operation as each $n-th$ sample depends on future samples as it is defined in the second term of the eqn. \ref{eq_imp}.

\subsection{SSA components: frequency domain}

The power spectrum of a times series is defined as the Fourier Transform of its autocorrelation function $r[m]= \mathbb{E}\{x[n]x[n+m]\}$

\[
S(e^{j\omega})=\sum_{m=-\infty}^{\infty} r[m]e^{-j \omega m}
\]

The autocorrelation function is an even function, $r[m]=r[-m]$ therefore the power spectrum is a real-valued function with positive values. In an SSA approach,  the entries of the matrix $\mathbf{R}$  represent estimates of the non-normalized  autocorrelation function, computed for $ m=-(M-1), \ldots, 0, \ldots, (M-1)$ while using the considered segment of data encompassing $N$ samples. The $m-th$ diagonal of the matrix has the values $r[m]$, where $m=0$ represents the main diagonal and  $ 1 \leq m \leq (M-1)$ are upper diagonals starting from the main diagonal.

However notice that the $r[m]$ in the different entries of the $m-th$ diagonal are numerically different, as  the time-delayed correlations, estimated with sub-segments of the data segment, are slightly different. This issue can be overcome if the correlation matrix  $\mathbf{R}$  is filled with the entries of the correlation function $r[m]$, instead of computing it with the embedding data matrix $\mb{X}$.

 Note that estimating the power spectrum with $2M-1$ samples of the autocorrelation function using a Discrete Fourier Transform (DFT), the resolution in frequency is the sampling rate divided by $2M-1$.


Furthermore, in linear invariant systems, the power spectrum of the output $S_{y}(e^{j \omega)}$ is related to the power spectrum of the input $S_x(e^{j \omega})$ by the transfer function $H(e^{j \omega})$ according to

\[
S_{y}(e^{j \omega)}= |H(e^{j \omega})|^2 S_x(e^{j \omega}).
\]


\subsubsection{Power Spectrum}
Denoting by $S_{min}$ and $S_{max}$ the extreme values of the power spectrum of the input signal $x[n]$, it is possible to show that the outputs of the analysis filters have energies $S_{min} < \mathbb{E}\{|y_m[n]|^2\} < S_{max}$. Computing the energy at the output of the filter $H_m(e^{j \omega})$ yields

\[
\frac{1}{2 \pi} \int_{0}^{2 \pi}S_x(e^{j \omega})|H_m(e^{j \omega})|^2 d \omega \leq S_{max} \frac{1}{2 \pi} \int_{0}^{2 \pi}|H_m(e^{j \omega})|^2 d \omega
\]

Then as the filter has unit energy, it follows from Parseval's theorem that
\[
\mathbf{u}^T_m \mathbf{u}_m = 1 \; \Rightarrow \; \frac{1}{2 \pi} \int_{0}^{2 \pi}|H_m(e^{j \omega})|^2 d \omega = 1.
\]

Hence, the upper bound is $S_{max}$. Similarly it is possible to show that energy is also lower bounded by $S_{min}$. Therefore, it can be concluded that the eigenvalues of the matrix $\mathbf{R}$ must be also bounded by  $S_{min} < \lambda_m < S_{max}$ as those values represent the energy of the extracted components. Remember that by eqn. \ref{eq_fil} the eigenvectors of the correlation matrix correspond to the filters. Then one can allocate the related eigenvalues to the frequency of the corresponding maximum of the frequency responses of the filters. In this way, the distribution of the eigenvalues matches the power spectrum of the input signal.

The synthesis filters show the same frequency profiles of the magnitude of the frequency response as the corresponding analysis filters. The  frequency range (passband) of both, the analysis and the synthesis filter, is identical, but the total energy of the equivalent cascaded filter $T_m(e^{j\omega})$ is different from the total energy of the individual filters. However, by computing the Fourier Transform of the equality eqn. \ref{eq_soma}, it follows that $\sum_m T_m(e^{j\omega}) = 1$. Therefore, the power spectrum of $ \hat{x}[n]$ is similar to the input if all components are summed up.
\begin{figure}
  \centering
  \includegraphics[width=0.47\textwidth]{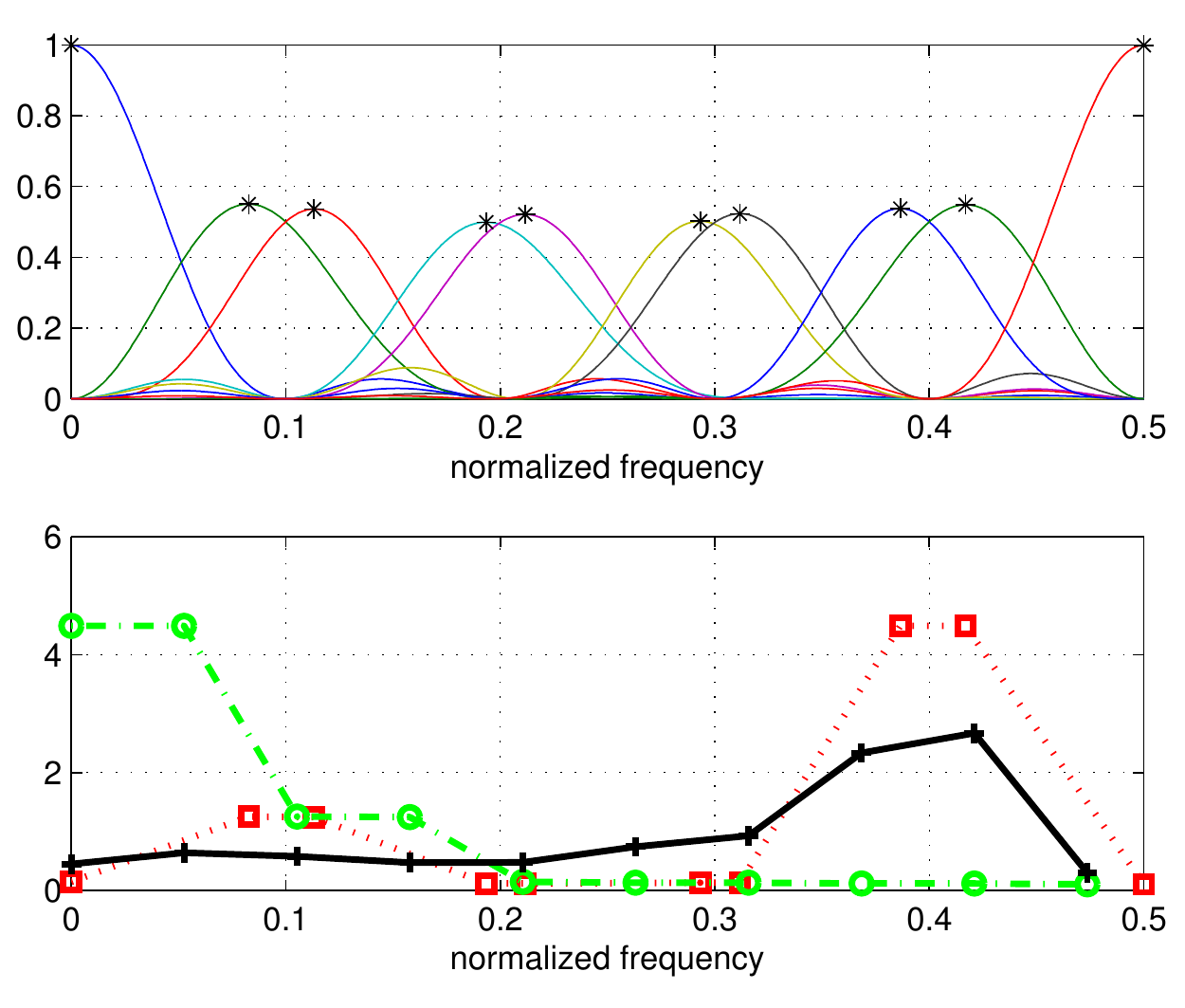}
   \includegraphics[width=0.5\textwidth]{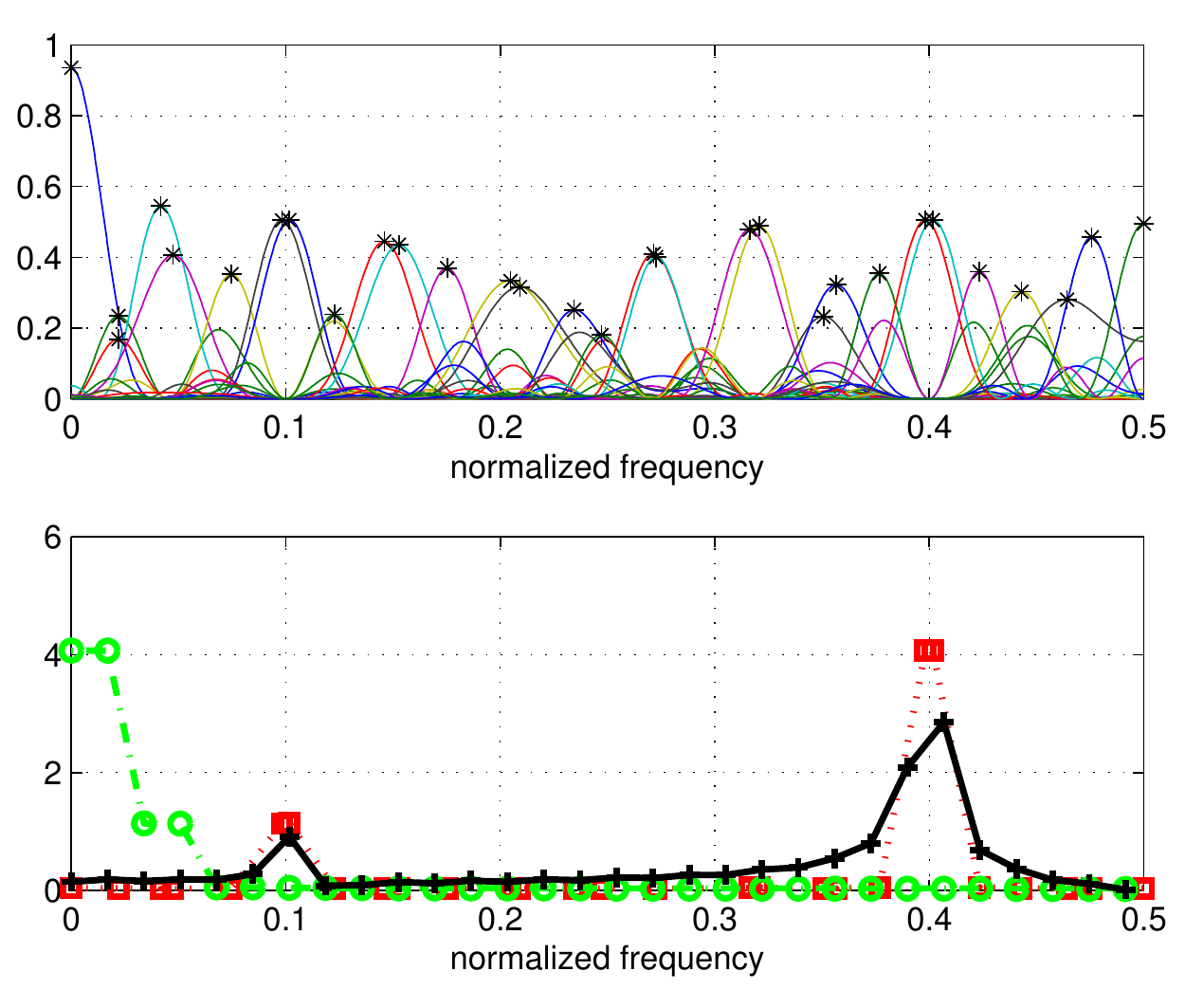}\\
  \caption{Illustrative example: SSA analysis.\emph{Left}:$M=10$ ; \emph{Right}: $M=30$. In both sides: \emph{Top}: frequency response of the filters and maximal values ($+$), \emph{Bottom}: scree plot of eigenvalues $o$, DFT of autocorrelation function ($+$); eigenvalue plot according to the localization of the maxima of the corresponding filters ($\square$).}
  \label{fignew}
\end{figure}

\paragraph{Illustrative Example}: A segment of the discrete signal
 \[
 x[n]=2 \sin(2 \pi (0.1)n)+4 \sin(2 \pi (0.4)n)+ \varepsilon[n],
  \]
  where $\varepsilon[n]$ represents Gaussian random noise (zero mean and unit variance), was analyzed with an SSA. Ideally, the power spectrum of the signal would consist of two pairs of delta Dirac functions at the frequencies of the sinusoids.  However, due to the use of a finite number of samples, the analysis shows the so-called spectral leakage. Furthermore, the number of samples determines the frequency resolution of the  estimated spectra. The numerical simulations were conducted for SSA with $M=10$ and $M=30$ and are compared with the estimation of the power spectrum using the DFT of the autocorrelation function $r[m]$ with $2M+1$ autocorrelation values. Fig.  \ref{fignew} illustrates the different results. On top, the frequency response of the $M$ filters are shown, and the maximal value of each filter is marked. On the bottom, the three plots correspond to the scree plot of the eigenvalues, and to the plots of the power spectrum estimates. Note that the scree plot shows four significant eigenvalues,
whatever is the value of $M$.  But also notice that for a scree plot, the values on the abscissa have no meaning. The power spectrum estimates correspond to the DFT of the autocorrelation function. Similarly, a power spectrum estimate is obtained by plotting the eigenvalues in accord with  the frequency of the maxima of the corresponding filters (eigenvectors). It has to be noticed that the frequency localization provided by the maximal values of the filters automatically identifies the frequencies of the related sinusoids. And note that, as the maxima of the filters are not equidistant, the localization can be more precise in cases when the frequency is not a multiple of the spectral resolution. The latter happens for the normalized frequency equal to $0.4$  as shown in the fig.  \ref{fignew}.

\section{Numerical Simulations}

The goal of the two following examples is
\bit
\item
to illustrate the application of SSA to gather relevant information from signals by
    \begin{itemize}
    \item
    extracting narrowband components from a wideband biomedical (EEG) signal or
    \item
    detecting the presence of a signal in communication (LTE) channels
    \end{itemize}
\item
to show that the frequency responses of the filters can be used to complement the eigenvalue spectrum in the conjugated Fourier domain.
\eit

\begin{figure}[!htb]
\begin{center}
  \includegraphics[width=0.48 \textwidth]{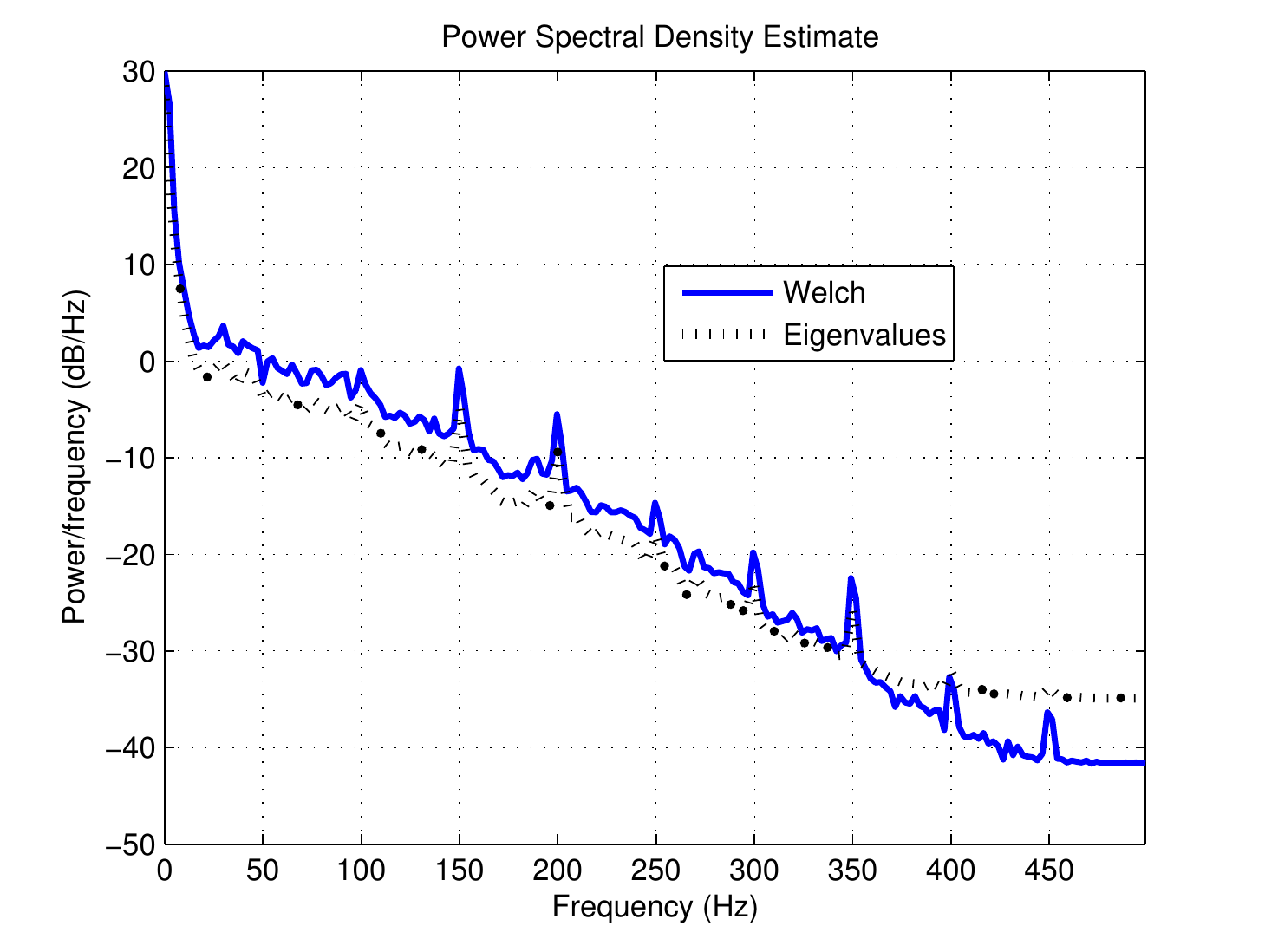}
   \includegraphics[width=0.48 \textwidth]{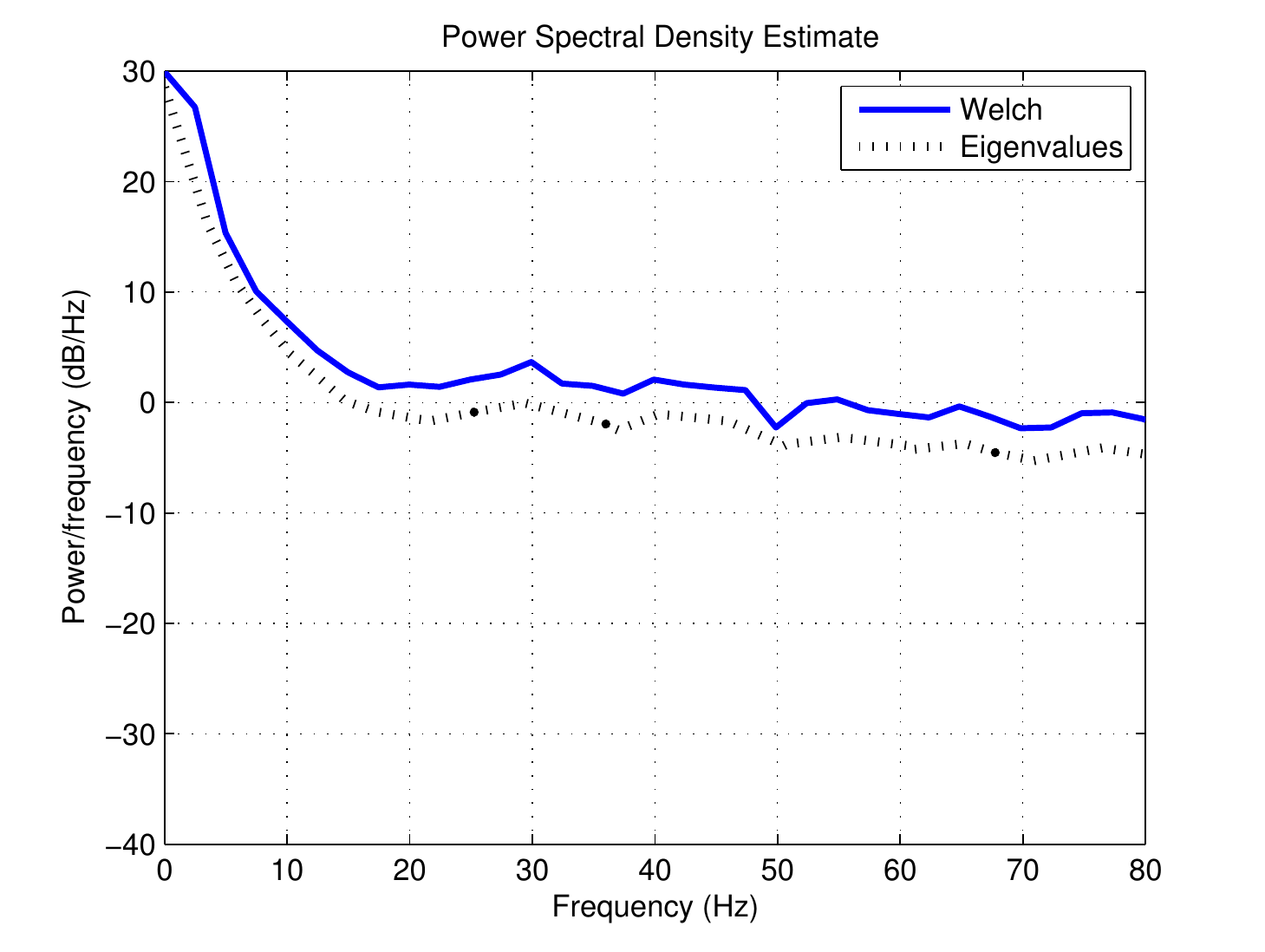}
\end{center}
  \caption{The power spectrum of the EEG signal, estimated either by Welch's method or represented by the eigenvalues  of the related correlation matrix: \emph{left}: $0-500 Hz$; \emph{right}: $0-80$ Hz}.
  \label{fig1}
\end{figure}

\begin{figure}[!htb]
\begin{center}
  \includegraphics[width=0.95 \textwidth]{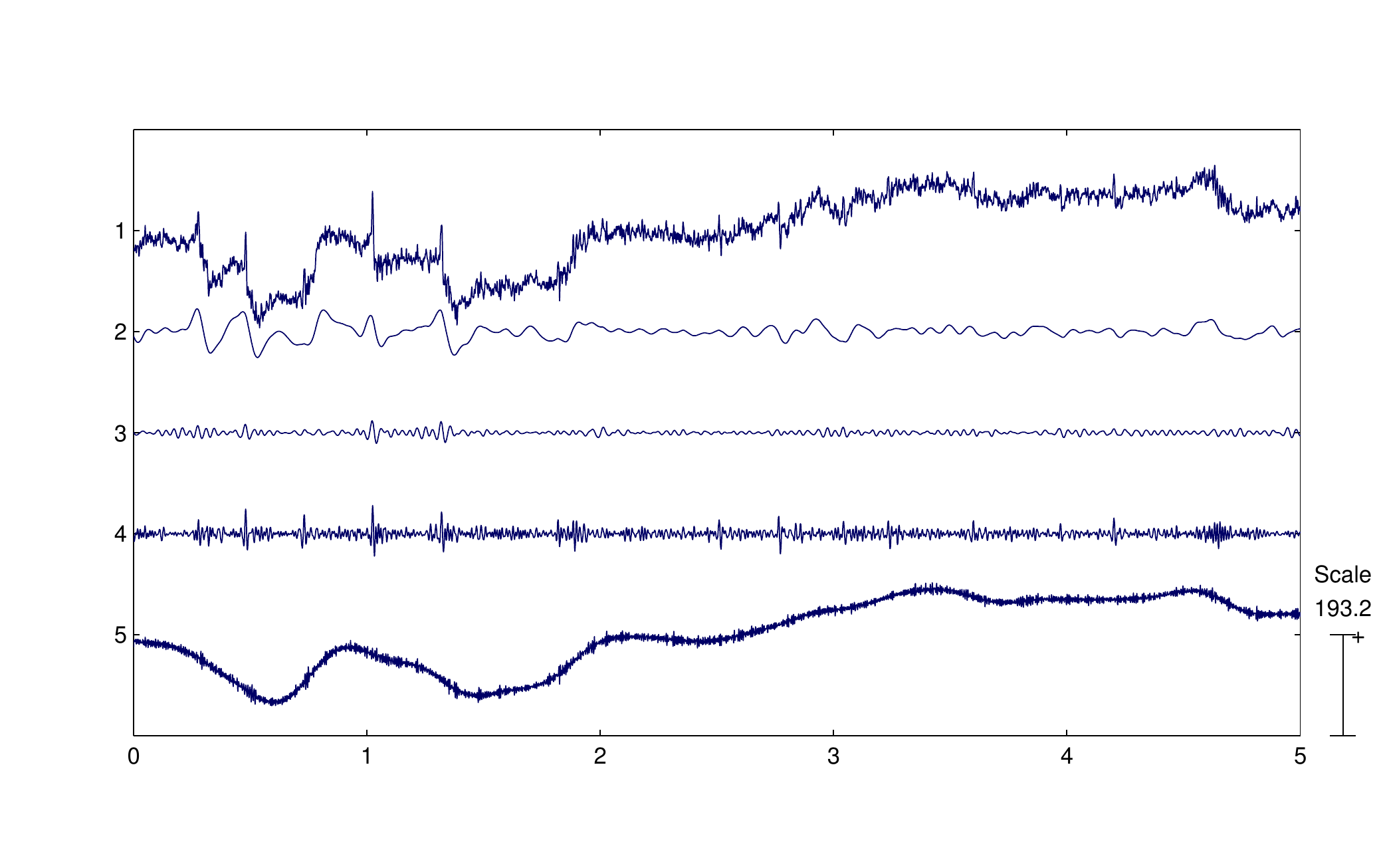}
\end{center}
   \caption{Original signal and four signal components, \emph{from top to bottom}:original, $\theta$ and $\alpha$  bands $2 - 15\ [Hz]$ ,the $\beta$-band ($15 - 25\ [Hz])$ and  $\gamma$-band ($25 - 100\ [Hz])$. Only $5\ [s]$ are shown for better visibility. }
   \label{fig2}
\end{figure}

\begin{figure}[!htb]
\begin{center}
    \includegraphics[width=0.48 \textwidth]{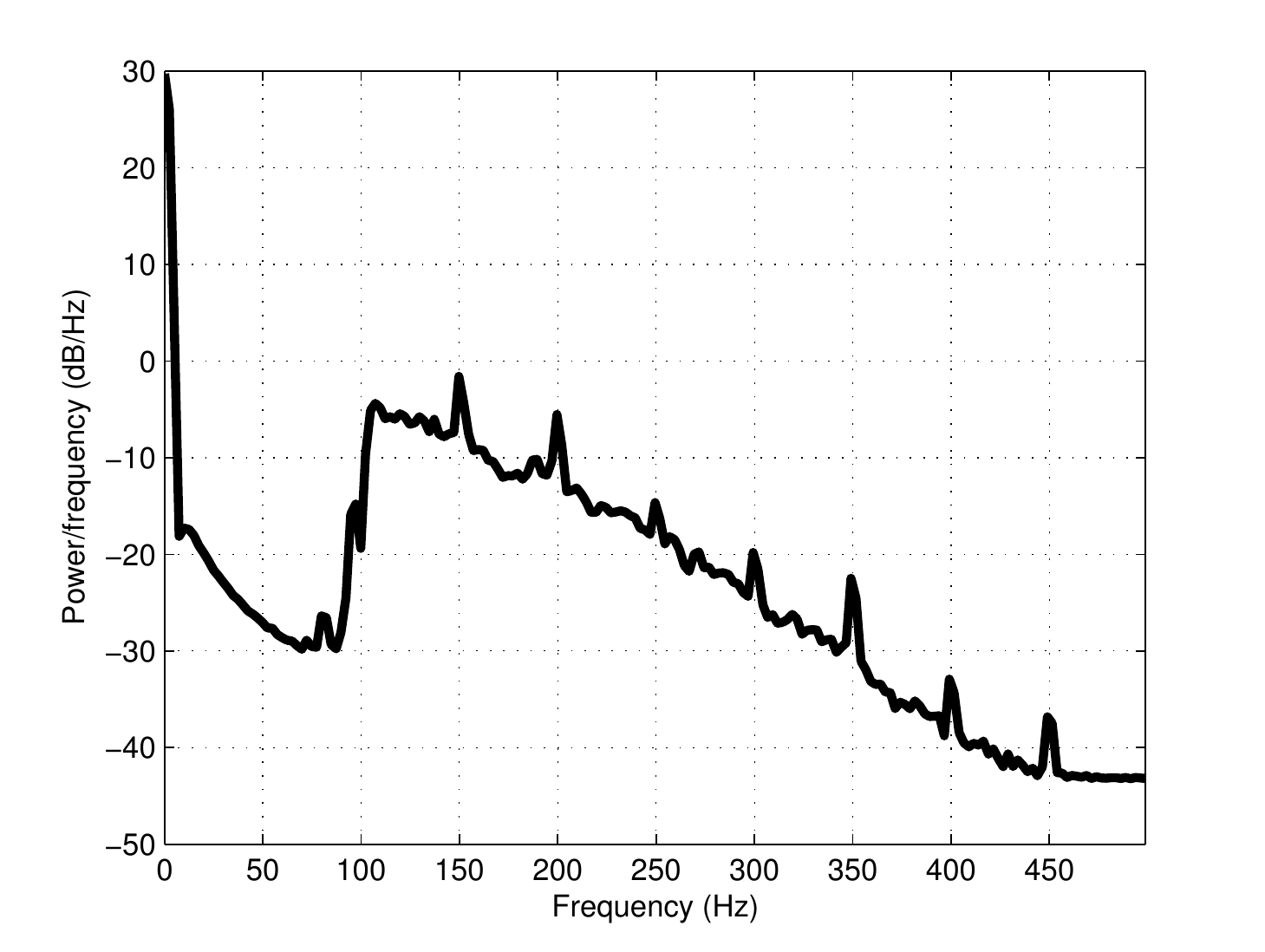}
   \includegraphics[width=0.48 \textwidth]{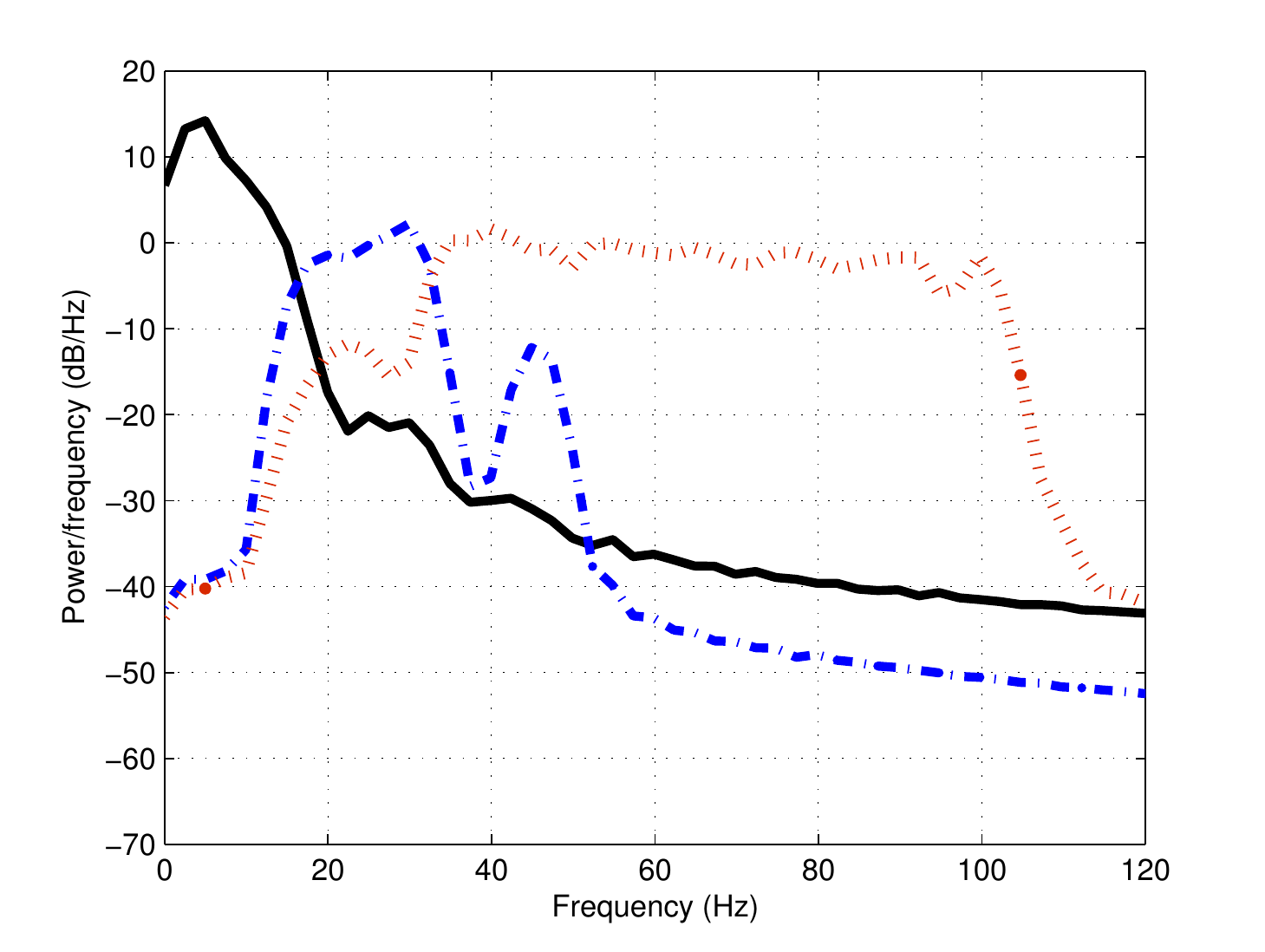}
\end{center}
   \caption{{\em Left}: The power spectrum of the fourth signal component shown in fig. \ref{fig2}. {\em Right}: An expanded view of the power spectra in the low frequency range of the narrowband $\theta$ - and $\alpha$ - bands ($2 - 15\ [Hz]$ - full line), the $\beta$-band ($15 - 25\ [Hz])$ - dash-dotted line) and $\gamma$-band ($25 - 100\ [Hz]$ - dotted line) are illustrated in fig. \ref{fig2}. All power spectra were computed using Welch's method.}
   \label{fig3}
\end{figure}
\subsection{Extraction of Narrow-Band Signals}

Typically in biomedical signal analysis, one acquires broadband signals but is interested only in some narrow-band signal components. As an example, figure \ref{fig1} shows the power spectrum of one segment of a single EEG channel recording, with a duration of $9\ [s]$ and a sampling rate of $1\ [kHz]$.  The power spectrum was estimated, employing Welch's method, in sub-segments of $401$ samples with $50\ \%$ of overlap. Figure \ref{fig1} clearly exhibits, outside of the band of interest, i.~e. $f > 100\ [Hz]$, interferences representing higher harmonics of the power line.  The same EEG segment was also analyzed with an SSA methodology, using an embedding dimension $M=201$. Next, the eigenvalues are plotted semi-logarithmically (see Fig. \ref{fig1}) versus frequency in the range $0 \le f \le 500\ [Hz]$.  The eigenvalues are ordered in accord with the corresponding maximum of frequency response of the  FIR filter. In this way, a spectral profile of the eigenvalues is generated which resembles the power spectrum
of the input signal. Note that the signal is decomposed into $M=201$ components in order to assure a frequency resolution close to $1000/401$.

In practice, one never is interested in all components resulting from the decomposition, rather only the sum of few components within certain frequency ranges are of interest mostly.
Fig. \ref{fig2} shows an example of signal components representing characteristics EEG bands. Here the top row signal represents the original signal, and the bottom trace shows signal components related to non-neuronal interferences like electro-oculograms, head/body movements and power line harmonics. The corresponding power spectra of the extracted signals are shown in figure \ref{fig3}. The signal components, resulting in traces $2 - 4$, are obtained by grouping together synthesized  components, i.~e. components that have been obtained after the last stage of the SSA processing chain as shown in fig. \ref{figblock}, in accord with the maxima of the frequency response of the corresponding FIR filter .

Therefore the frequencies of the  maxima of the filters are used to group the components $\hat{x}_m [n]$ into  four clusters: three related with different characteristic EEG bands and the fourth is considered noise related. Then, the components forming one cluster are added up to result in a narrow band signal containing its frequency contents in one of the characteristic EEG bands. More specifically, referring to Fig. \ref{fig2}, six SSA components were combined to result in traces 2 and 3, each, while trace 4 encompasses $28$ SSA components and , finally, trace 5 represents a combination of the remaining $162$ SSA components. The related power spectra are show in Fig. \ref{fig3}.

\subsection{Detection of radio frequency signals}

\begin{figure}[!htb]
\begin{center}
\includegraphics[width=90mm,height=45mm]{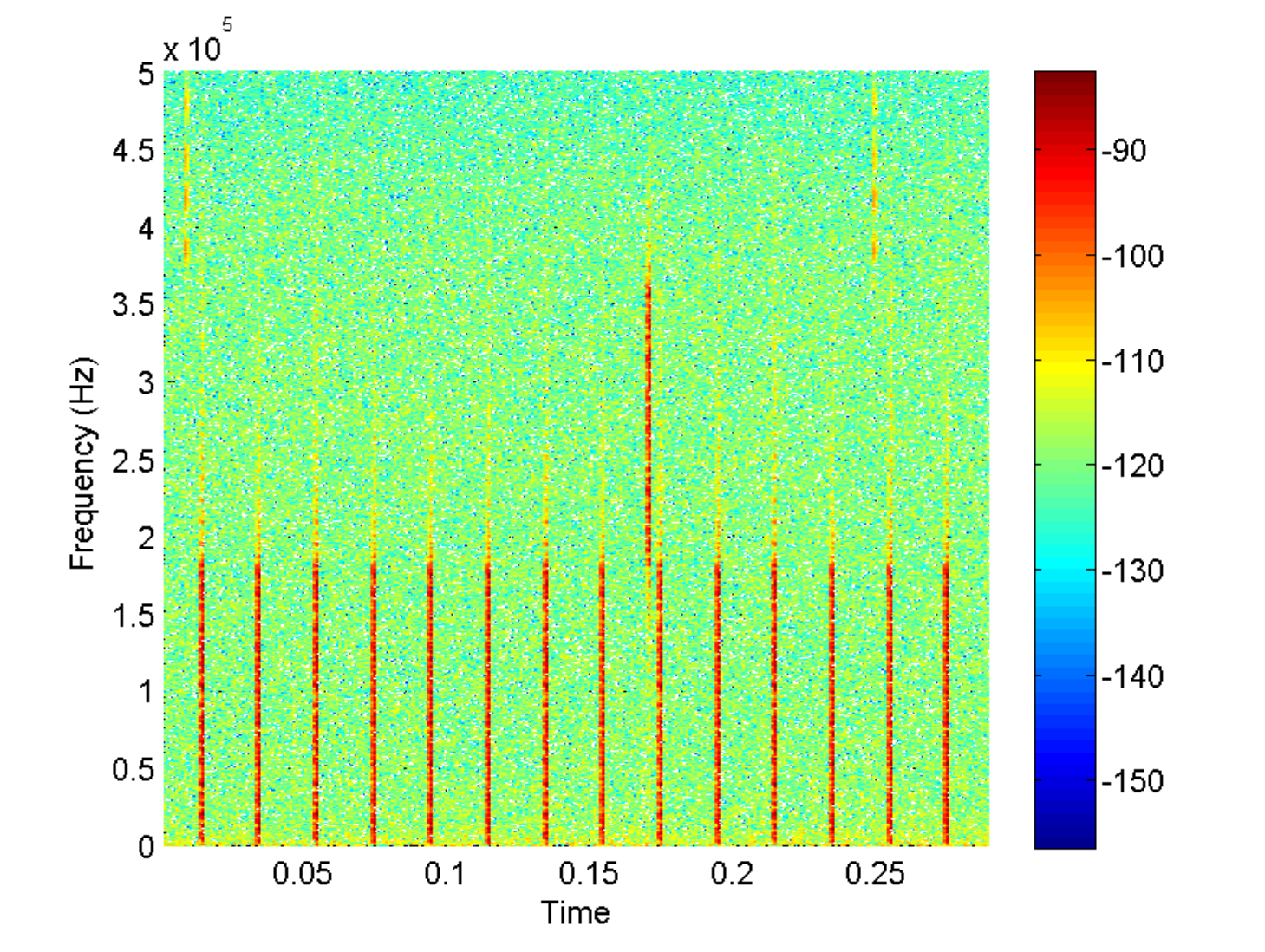}
\end{center}
 \caption{Spectrogram of a Long Term Evolution (LTE) signal as described in the text.}
\label{fig4}
\end{figure}
The next example is related with a universal mobile telecommunication system (UMTS) signal, more specifically a signal from the Physical Uplink Control Channel (PUCCH) of the Long Term Evolution (LTE) uplink \citep{Perez2015}.  The data was acquired with an USRP B200 Software Defined Radio board \footnote{https://www.ettus.com/product/details/UB200-KIT}, at a sampling rate of $1~[MHz]$ and with a central frequency of  $2.551\ [GHz]$. Figure \ref{fig4} illustrates the spectrogram of the acquired signal from DC to $500\ [kHz]$. It can be verified that the  signals are present in short segments of time ($0.5\ [ms]$). In this example, the LTE signal comprises three resource blocks (RB) of the LTE,  each with bandwidth of $180\ [Khz]$ which are assigned to PUCCH channels. The PUCCH is only occupied when a User Equipment (UE) needs to send control information.

\begin{figure}[!htb]
\begin{center}
\includegraphics[width=90mm,height=45mm]{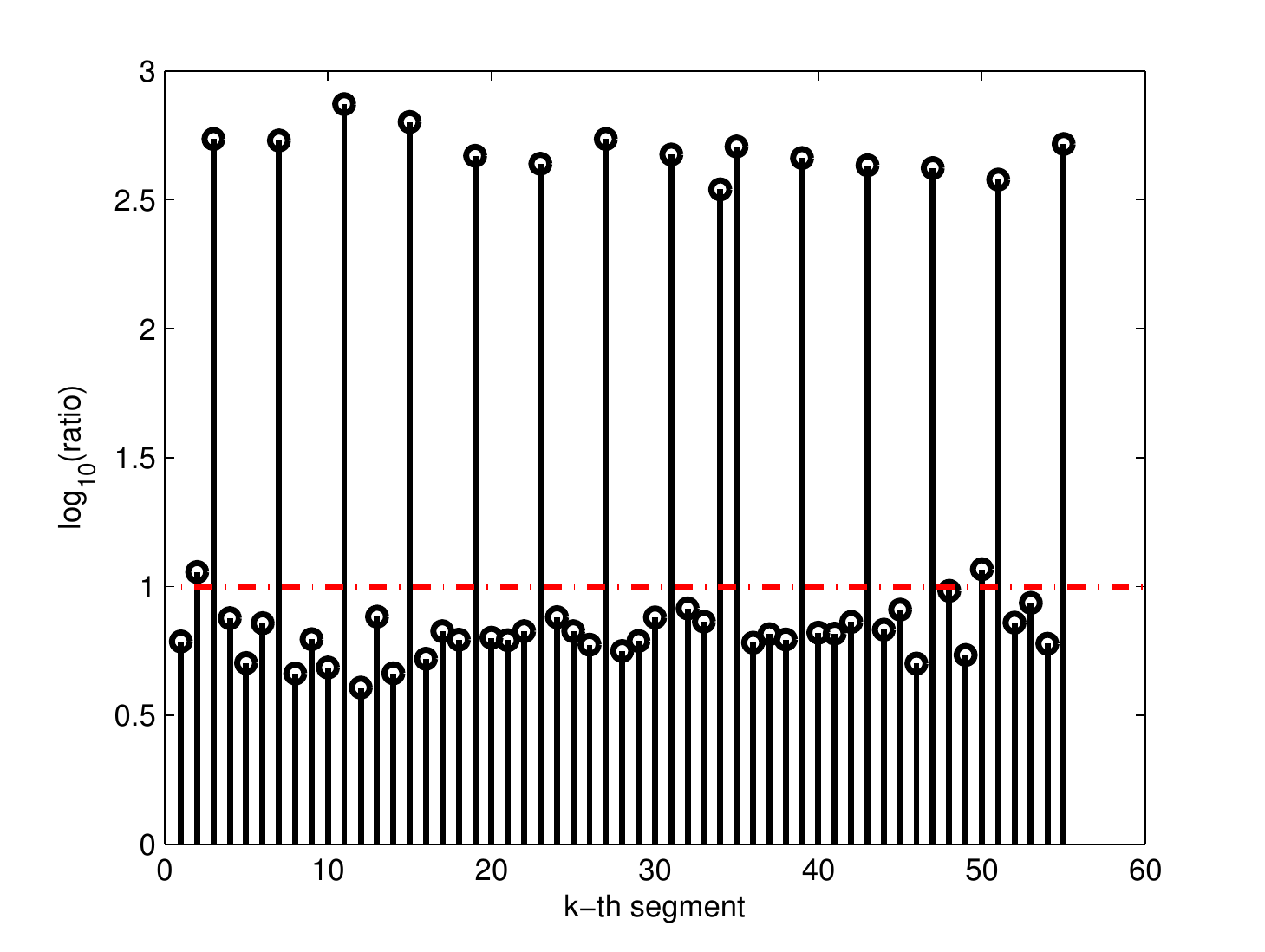}
\end{center}
\caption{Eigenvalue ratios of the $55$ segments of the LTE signal}
\label{fig5}
\end{figure}

\begin{figure}[!h]
\begin{center}
\includegraphics[width=130mm,height=80mm]{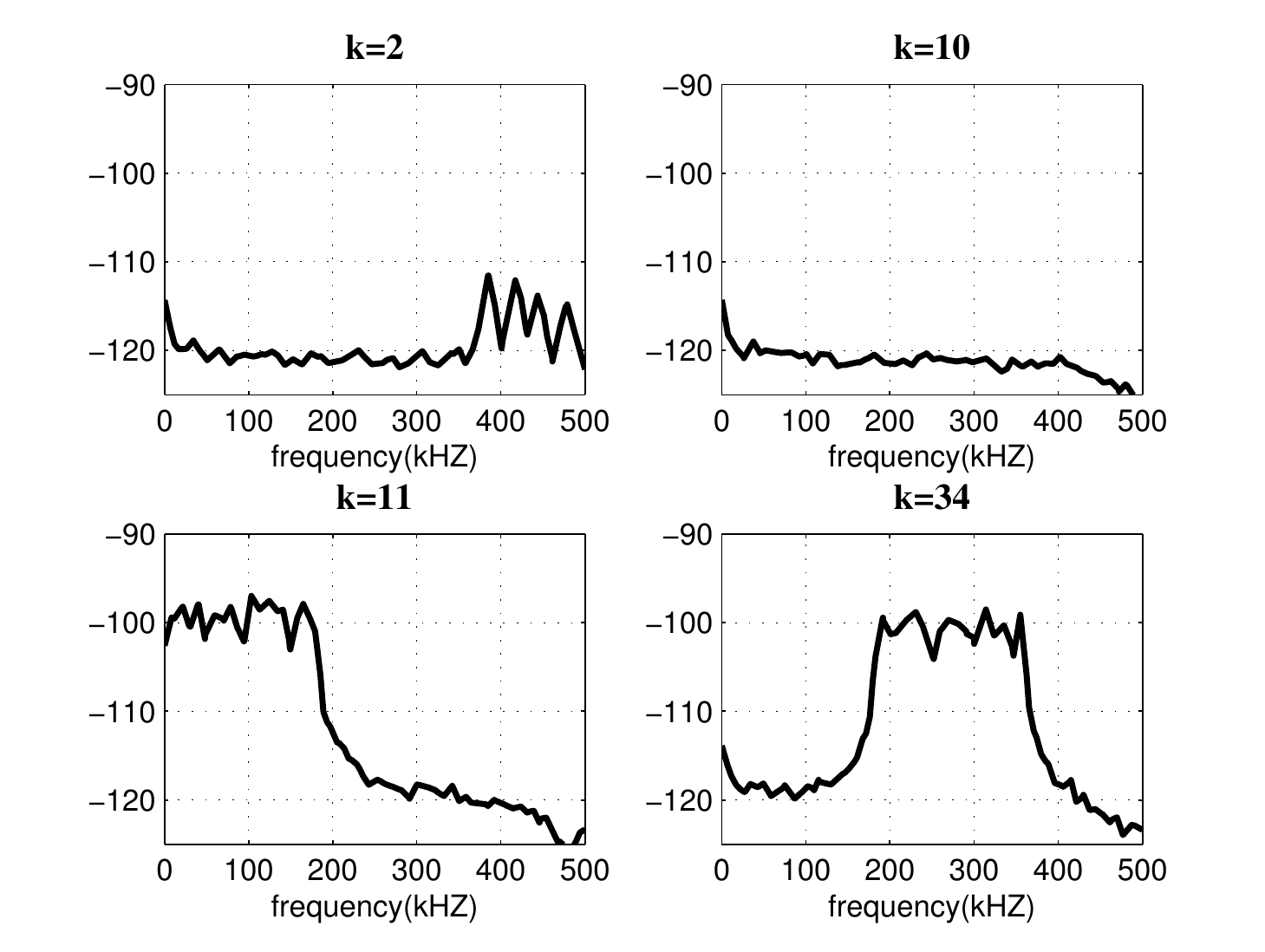}
\end{center}
\caption{Semi-logarithmic eigenvalue plot of $4$ segments of $N=5000$ samples. The unit on the ordinate is $dB/Hz$.  }
\label{fig6}
\end{figure}

For each segment of $N=5000$ samples of the SSA model with an embedding dimension $M=100$, an eigenvalue ratio is estimated for the $55$ sub-segments of the signal shown in the spectrogram. The figure \ref{fig5} shows the eigenvalue ratio, in semi-logarithm scale, for all sub-segments $k=1, 2, \ldots 55$. It can be concluded that it is possible to define a threshold to decide if a signal is present or not, in each of the sub-segments. However, note that with this detection it is not possible to know which of the bands is occupied. However if the eigenvalues are plotted according to the frequency of the maxima of the frequency response of the corresponding filters the frequency localization is possible. Figure \ref{fig6} shows that in the second segment ($k=2$) the low power signal occupies the highest frequency band ($f > 360\ [kHz]$). The figure also shows the eigenvalue spectrum when there is no signal ($k=10$). The next two plots  concern the occupancy of the low-pass band and the band-pass, respectively.

\section{Conclusions}
The interpretation of singular spectrum analysis (SSA) as a bank of filters can be useful to attain a more clear-cut insight into the outcomes of the method. By the frequency responses of the filter bank, corresponding to the eigenvectors of the correlation matrix, the frequency content of the different component signals can be easily attained. SSA filters are data-adaptive, and the relevance of one component to the energy of the input signal is deduced from the corresponding eigenvalue. In this study, the relation between the eigenvalues and the power spectrum was addressed. It was shown that it is advantageous to re-order the eigenvalues in accord with the occurrence of the maxima of the  frequency responses of the related filters rather than simply arranging them by decreasing value as is common practice. In summary, exploiting the information from both analysis strategies, i.~e. the eigenvalue spectrum and the related frequency response of the filter, provides additional insight compared to a canonical analysis.

Two typical applications illustrated the advantage of this analysis technique. Analyzing a typical biomedical signal, more specifically an EEG, often amounts to extracting narrow-band signal components from a broadband acquired signal. It is known, and corroborated in fig. \ref{fig1}  that the power spectra of most biomedical signals show a $1/f$ spectral distribution \citep{Li2012}, which naturally results in a standard ordering of the eigenvalues by decreasing value. But the peaks, belonging to the higher order harmonics of the line noise spectral contribution, can only be seen if the eigenvalues are properly matched to the maxima of the frequency responses of the filters. Thus, to assure that the extracted components fall in a chosen narrowband range, namely the classical EEG frequency bands, the  frequency profiles of the related filters have to be studied to identify those signal components which need to be combined to result in the desired narrowband signals. A second prototypical application is channel identification in spectrum sensing. The goal here was to identify at times unoccupied transmission channels which then can be used for additional signal transmission. These simple applications illustrate the usefulness of a combined analysis of the eigenspectrum of the correlation matrix in the time domain as well as the related frequency response. It was shown that re-ordering the eigenvalues in accord with the maxima of the frequency responses of the filters allows to identify the most {\em relevant} signal components for the problem at hand.

\section*{Acknowledgements}
This work was supported by FCT:  grant
 to A. R. Teixeira (Ref:SFRH/BPD/101112/2014)
and the project EXCL/EEI-TEL/0067/2012 (CREaTION).


\section*{Appendix}
SSA MATLAB implementation and two of the examples of the report.
\begin{verbatim}
a=[2 4]
f=[0.1 0.4]
x=zeros(1,1024)
n=0:length(x)-1;
for k=1:2
    x=x+a(k)*sin(2*pi*f(k)*n);
end
x=x+randn(1,length(x));
M=9
%Correlation matrix symmetric
model=model=SSAglobal(x,M,'toeplitz','eigenvectors')
%Correlation matrix non-symmetric
model1=SSAglobal(x,M,'embedding','eigenvectors')
% Grouping components according to model.peaks
% EEG sampled 1000Hz,embedding dimension 201
%model= SSAglobal(s,200,'toeplitz','eigenvectors',1000)
F=[2 8;8 15;15  32; 32 100]
 L=size(F,1)
% Components
 group=zeros(L+1,9000)
 contador=zeros(1,L+1);

   for c=1:L
        a= (model.peaks>= F(c,1) & model.peaks <F(c,2))';
        switch sum(a)
        case 0 % no peak
          group(c,:)=zeros(1,M)
        case 1
           group(c,:)=model.comp(a,:);
        otherwise
           group(c,:)=sum(model.comp(a,:));
        end
        contador(c)=sum(a);
   end
   %the
   a=(model.peaks< F(1,1) | model.peaks>F(L,2))';
   contador(L+1)=sum(a)
   group(L+1,:)=sum(model.comp(a,:))
%%%

%%%%%%

function [model]=SSAglobal(x,M,TYPE,ORDERING,Fs)
% model=SSAglobal(x,M,TYPE,ORDERING,Fs)
% model=SSAglobal(x,M)
%  Default: TYPE='toeplitz',ORDERING='eigenvalues',Fs=1 (normalized
%  frequency--radians)
%model=SSAglobal(x,M,'toeplitz','eigenvectors')
%OUTPUT:
% model.eigval : eigenvalues (ordering by decreasing order or according to
%                the peaks of TF)
% model.all: filter bank (analysis and synthesis)
% model.TF: frequency responses of the analysis-synthesis
% model.peaks: frequency of the peaks (if Fs=1, 0<values<0.5)
% model.f:   f=(0:N-1)*(Fs/N);
% model.comp : components outputs of all filters of model.all
% AMT October 2015


N=4000;   %N

if nargin <2 | nargin>5
    error ('invalid number of arguments');
else
    switch nargin
        case 2
            TYPE='toeplitz';
            ORDERING='eigenvalues';
            Fs=1;
        case 3
             ORDERING='eigenvalues';
             Fs=1;
        case 4
            Fs=1;
    end
end

 if strcmp(TYPE,'toeplitz')
%     %compute autocorrelation function (from delay -M up to M)
    rx=xcorr(double(x),M);
    R=toeplitz(rx(M+1:end));

 elseif strcmp(TYPE,'embedding')
    fim=0;
    for k=1:M+1
        X(k,:)=x(M+1-k+1:end-fim);
        fim=fim+1;
    end
    R=X*X';
else
    error('valid options: embedding or toeplitz');
 end

% order of eigenvalues by decreasing
[U,S,dummy]=svd(R);
% %
% % filter bank
 for k=1:M+1
     u=U(:,k);
     %normalization
   t(k,:)=conv(u,u(end:-1:1))./(M+1);
 end
%
% %%%ORDERING BY eigenvalues (default)
%  %filter bank  and eigenvalues by dcreasing order
  model.all=t;
  model.eigval=diag(S);
  model.U=U;
%
 f=(0:N-1)*(Fs/N);
 Tf=abs(fft(t,N,2));
%  %Only half... 0 to Fs/2
 Tf(:,N/2+2:end)=[];
 model.TF=Tf;
 model.peaks=[];
 model.f=f(1:size(Tf,2));
if strcmp(ORDERING,'eigenvectors')
     [dummy,inx]=max(Tf,[],2);
     [dummy,inx]=sort(f(inx));
     model.eigval=model.eigval(inx);
     model.U=U(:,inx);
     model.all=model.all(inx,:);
     model.TF=Tf(inx,:);
     model.peaks=dummy;
end
%%The components
for k=1:M+1
    y=conv(model.all(k,:),x);
    model.comp(k,:)=y(M+1:end-M);
end

\end{verbatim}


\end{document}